\documentstyle[twocolumn,prl,aps]{revtex}
\begin{document}
\draft
\wideabs{
\title{The Josephson Plasma Resonance in $\rm Bi_2Sr_2CaCu_2O_8$\\in a Tilted Field.}
\author{Sibel P. Bayrakci, Ophelia K.\,C. Tsui\cite{byline}, and N.\,P. Ong}
\address{Joseph Henry Laboratories of Physics, Princeton University, Princeton, New 
Jersey 08544}
\author{K. Kishio, S. Watauchi}
\address{Department of Applied Chemistry, University of Tokyo, Hongo, Bunkyo-ku, Tokyo 
113, Japan}

\date{\today}
\maketitle
\begin{abstract}
The dependence of the Josephson plasma frequency $\omega_p$ in $\rm Bi_2Sr_2CaCu_2O_8$ on 
a tilted field $\bf H$ is reported.  Measurements over a large range of $B$ and tilt 
angle $\theta$ allow a detailed comparison with a recent calculation by Koshelev.  With a 
slight modification of the model, close agreement is obtained.  From the fits, we find 
values for the in-plane correlation length and the zero-field critical current density 
$J_{c0}$ (4,600 A/cm$^2$ at 30 K).  An analogy to Bragg diffraction is described, as well 
as a picture for the fractional-exponent behavior of $\omega_p$ versus $\bf H$.
\end{abstract}
\bigskip
}
In cuprate superconductors with large coherence-length anisotropies ($\gamma > 10^2$), 
the Josephson coupling energy between adjacent layers (or bilayers) is quite small.  The 
Josephson plasma frequency $\omega_p$, which measures the coupling between layers, has 
been observed in $c$-axis reflectivity experiments at far-infrared frequencies 
($\omega_p/2\pi \sim 30-70 {\rm cm}^{-1}$ in the single-layer cuprate $\rm La_{2-
x}Sr_xCuO_4$) \cite{Uchida}.  In the bilayer cuprate $\rm Bi_2Sr_2CaCu_2O_8$ (Bi 2212), 
the large separation $s$ between adjacent bilayers ($s \sim 15.8 \AA$) weakens the 
Josephson coupling significantly.  An applied magnetic field $\bf H$ further suppresses 
$\omega_p$ to frequencies in the microwave range (10-100 GHz; 1 cm$^{-1}\simeq$ 30 GHz).  
The plasma resonance in Bi 2212 is readily observed as sharp resonances in the absorption 
of incident microwave power as $H$ is slowly swept 
\cite{Tsui1,spie,Yuji1,Tsui2,Yuji2,Maeda}.  Empirically, the field dependence of 
$\omega_p$ may be described by $\omega_p^2 \sim H^{-\mu}$, with $\mu \sim 0.8$ when $\bf 
H\parallel c$ \cite{Tsui1,spie,Tsui2}.  

Recently, the identification of the resonant absorption with excitation of the plasma 
mode has been called into question by Sonin \cite{Sonin}, who discusses a number of 
difficulties with the plasma mode interpretation \cite{spie,Yuji1,Tsui2,bmt,Bula1}.  To 
help distinguish between competing models, quantitative comparisons between experiment 
and theory seem desirable.  The strong variation of the resonance field with field tilt 
angle $\theta$ (the angle between $\bf H$ and the $ab$ plane) is a highly specific 
feature of this phenomenon.  We report measurements over a wide range of fields and 
$\theta$, and compare the data with a recent model by Koshelev \cite{Koshelev}.

We briefly review the Josephson plasma mode.  The transfer of $n_p$ Cooper pairs between 
two Josephson-coupled superconducting layers leads to a charging energy cost of 
$(2n_pe)^2/2C$, where $e$ is the elementary charge, and $C$ the unit-area capacitance.  
In addition, deviation of the relative phase $\delta\varphi$ from zero incurs a cost in 
energy $E_J(1-\cos\delta\varphi)$ where $E_J$ is the Josephson energy (per unit area).  
Hence, the Hamiltonian with $n_p$ and $\delta\varphi$ as conjugate variables is (for $\bf 
H$ = 0) \cite{Anderson}
\begin{equation}
{\cal H_J} = (2n_pe)^2/2C + E_J (1-\cos\delta\varphi).
\label{pendulum}
\end{equation}
In analogy with the pendulum, the charging energy oscillates out-of-phase with the 
Josephson energy at the resonance frequency $\omega_p$ given by $\omega_p^2 = 
(2\pi/\phi_0)^2E_J/C$ where $\phi_0 = h/2e$ is the flux quantum ($h$ is Planck's 
constant).  In finite field, pancake vortices in the layers strongly distort the local 
phase by adding a component $\varphi_n({\bf r})$ that is static on the time scale of 
$\omega_p$ (where ${\bf r} = (x,y)$ and $n$ is the layer index; we choose axes with $\bf 
c \parallel \hat{z}$ and $\bf H\times c \parallel \hat{x}$).  If the vortex array is 
highly disordered, $\varphi_n({\bf r})$ fluctuates stochastically with $\bf r$ and $n$.  
Adding this static phase to the dynamic phase variable in Eq.\,\ref{pendulum}, we find 
that $\omega_p^2$ is reduced \cite{bmt,Bula1} by the factor $ \langle\cos 
\phi_{n,n+1}\rangle $ (thermally averaged and integrated over the sample), where 
\begin{equation}
\phi_{n,n+1}({\bf r}) \equiv \varphi_{n+1}({\bf r})-\varphi_n({\bf r})- 
(2\pi/\phi_0)\int^{n+1}_n d{\bf \ell \cdot A}
\label{phi}
\end{equation}
is the gauge-invariant phase difference between layers $n$ and $n+1$ with $d{\bf 
\ell\parallel c}$, and $\bf A$ is the vector potential.  The factor $\langle\cos 
\phi_{n,n+1}\rangle$ contains the main effect of field on $\omega_p$.  In general, 
quantitative comparisons of calculations of $\langle\cos \phi_{n,n+1}\rangle$ with 
experiment are difficult because they require detailed knowledge of vortex 
configurations. 

If the Josephson energy is small, however, the vortex positions at high temperatures are 
very weakly correlated between adjacent layers.  In the presence of a tilted field, the 
flux density parallel to the layers is nearly uniform and close to the applied in-plane 
field induction $B_y$.  While the vortices in each layer produced by $B_z$ remain highly 
disordered, the in-plane component couples to $\phi_{n,n+1}$ {\em uniformly} throughout 
the sample.  This situation provides a more transparent comparison with experiment.  

First, we assume $\bf H\parallel c$.  In the high-temperature limit, complete randomness 
in vortex positions should suppress $ \langle\cos \phi_{n,n+1}\rangle $ to a value close 
to zero.  As $T$ decreases, we expect correlations of the vortex positions to increase 
(within each layer and between layers).  Clearly, the magnitude of $ \langle\cos 
\phi_{n,n+1}\rangle $ grows in proportion to the strength of these correlations.  The 
relation between $ \langle\cos \phi_{n,n+1}\rangle $ and the phase correlation function 
$\langle \exp[i (\phi_{n,n+1}({\bf r})-\phi_{n,n+1}(0))] \rangle_0$ is derived by 
expanding the partition function in the small parameter $\beta E_Ja^2$ (where $\beta = 
1/k_BT$ and $a^2 = \phi_0/B_z$) \cite{Koshelev}.  We find 
\begin{equation}
\langle\cos \phi_{n,n+1}\rangle \simeq \frac{\beta E_J}{2}\int d^2r 
\langle \exp[i (\phi_{n,n+1}({\bf r})-\phi_{n,n+1}(0))] \rangle_0.
\label{correlate}
\end{equation}
As $\omega_p^2\sim\langle\cos \phi_{n,n+1}\rangle$, Eq.\,\ref{correlate} states that the 
plasma frequency (squared) is the average of the phase correlation function over the 
sample.  By the definition of $\phi_{n,n+1}$, it is clear that the correlation involves 
$\varphi_n$ at 4 points (2 points separated by $\bf r$ in each layer).  If in-plane 
correlation dominates the interlayer correlation, the integrand in Eq.\,\ref{correlate} 
reduces to the modulus-squared of the {\em single}-layer correlation function $S_1({\bf 
r}) = \langle \exp [i(\varphi({\bf r})-\varphi(0))] \rangle_0$ \cite{Koshelev}.  

Next, we consider tilting the field.  If $B_y$ is assumed to be uniform, it simply adds 
to the phase difference in Eq.\,\ref{correlate} a term proportional to the flux threading 
the area defined by the 4 points, viz. $(2\pi/\phi_0)B_ysx$ [the $\bf A$ term in 
Eq.\,\ref{phi}].  Equation \ref{correlate} then simplifies to 
\begin{equation}
\langle\cos \phi_{n,n+1}\rangle \simeq \frac{\beta E_J}{2}\int d^2r |S_1({\bf r})|^2 {\rm 
e}^{-iqx},
\label{Fourier}
\end{equation}
where $q= 2\pi sB_y/\phi_0$.  We have the key result that $B_y$ adds a phase shift that 
increases linearly with $x$ with the wavevector $q$.  Upon integrating over the volume, 
we obtain the Fourier transform of $|S_1({\bf r})|^2$, which may be observed as 
$\omega_p^2$ versus $B_y$.  Thus, in a tilted field, information on the in-plane 
correlation of the vortices, inserted by $B_z$, is obtained by measuring $\omega_p$ 
versus $B_y$.

It is instructive to compare this situation with the behavior of an ideal, low-$T_c$ 
junction in a field.  There, the Josephson coupling is uniform over the length $L$ of the 
junction.  The field produces a Fresnel interference pattern with period fixed by $L$, 
identical to that produced by the diffraction of light through a slit of uniform 
transparency.  In the present case, however, vortex disorder strongly modulates the 
Josephson coupling on some correlation length-scale.  The Fourier pattern produced is 
closer in analogy to Bragg diffraction from a liquid.  The predicted behavior of 
$\omega_p^2$ versus 1/$\theta$ is analogous to the variation of the structure factor 
${\cal S}(\bf q)$ versus $q$ measured in a liquid with strong correlation.  From the 
relation $qa = 2\pi s (B_y/\phi_0 \tan\theta)^{1/2}$, we find that the small $qa$ regime 
is explored in a weak $\bf H$ tilted away from the plane, while the large-$qa$ range is 
accessed with an intense $\bf H$ with $\theta\rightarrow 0$.

\mbox{Finally, with the approximation $|S_1({\bf r})|^2 \sim \exp(-r^2/\xi^2)$,} we have 
Koshelev's result \cite{Koshelev}
\begin{equation}
\langle\cos \phi_{n,n+1}\rangle \simeq \frac{\beta E_Jfa^2}{2}\exp\left[-f \frac{\pi 
s^2B_y^2}{\phi_0B_z}\right],
\label{Koshelev}
\end{equation}
where $\xi$ is the in-plane correlation length, and  $f \equiv a^{-2}\int d^2r |S_1({\bf 
r})|^2 = \pi \xi^2/a^2$ equals the number of vortex pancakes within a correlated patch.  
We compare Eq.\,\ref{Koshelev} with our experiment.

Crystals of $\rm Bi_2Sr_2CaCu_2O_8$ with transition temperature $T_c\sim $ 92 K were cut 
from a boule grown in an image furnace and annealed in flowing oxygen to obtain optimum 
oxygen content.  The samples were glued to a diamond substrate and placed
in an oversized circular waveguide.  Excitation of the plasma mode was detected 
bolometrically as sharp resonances in the absorption of microwave power versus field.  
The absorption was detected by a cernox sensor. 

At each tilt angle $\theta$, the resonance field $B_0$ is determined by slowly sweeping 
the field at fixed microwave frequency $\omega$ \cite{Tsui1,Tsui2}. The inset in 
Fig.\,\ref{fig1} displays sweep-up and sweep-down traces at 50 K and 48 GHz, for three 
values of $\theta$.  By repeating the sweep at 4 to 6 values of $\omega$ (between 30 and 
70 GHz), we map out the dependence of $\omega_p$ on $B$ at that $\theta$.  The procedure 
is repeated at $\sim 20 $ values of $\theta$ to generate the full dependence of $B_0$ on 
$\omega$ and $\theta$.  The main panel of Fig.\,\ref{fig1} shows the variation of $B_0$ 
with $\theta$ at five values of $\omega$. With decreasing $\theta$, $B_0$ rises to a 
maximum value, and then decreases linearly to a cusp at alignment. In Fig.\,\ref{fig2}, 
we display a different cut of the data by plotting $\omega$ versus $B_0$ at fixed 
$\theta$ in log-log plot.  The solid lines in both figures are fits to a slightly 
modified form of Koshelev's expression which we discuss below. 

We summarize the key features observed. Over our range of $\omega$, it is convenient to 
approximate the field variation of $\omega_p$ by the power-law, $\omega^2 = AB_0^{-\mu}$, 
where $A$ and $\mu$ are empirical parameters that depend on $\theta$ and $T$.  When $\bf 
H$ is normal to the layers, the power-law exponent of $\omega_p^2$ vs $H$ is {\it 
fractional}, viz. $\mu \sim$ 0.8.  For small tilts of $\bf H$ away from the $c$-axis, the 
power law with the same $\mu$ and $A$ remains valid, if we replace $H$ with $H_z$, as 
expected for a highly anisotropic system. However, as $\bf H$ is tilted closer to the 
plane ($\theta< 15^\circ$), strong deviations from this simple scaling become apparent.  
While the data may still be described by a power law, the exponent $\mu$ rises rapidly to 
1.80 as $\theta\rightarrow 0$. The plasma frequency $\omega_p$, observed at fixed $H$, 
increases rapidly to a maximum at $\theta$ near 1$^\circ$, and then decreases linearly 
with $|\theta|$ to a finite value as $\theta\rightarrow 0$.  For experimental reasons, it 
is convenient to plot $B_0$ versus $\theta$ at fixed $\omega$.  Nonetheless, the same 
trend is observed.  $B_0$ rises to a maximum at $\theta\sim 1^\circ$, and then decreases 
to a finite value at $\theta=0$.  In this narrow range of $\theta$, the empirical 
parameters $\mu$ and $A$ change rapidly, reflecting the increasing influence of the 
Josephson vortices.

As it stands, Koshelev's expression Eq.\,\ref{Koshelev} is in close agreement with the 
observed dependence of $B_0$ on $\theta$, if $\omega$ {\em is kept fixed} while $\theta$ 
is varied (because $\omega$ is the independent experimental variable, we need to invert 
Eq.\,\ref{Koshelev} to calculate the $B$ versus $\theta$ profile). Fits to the measured 
$B_0$ at angles $\theta > 1^\circ$ are shown as solid lines in the main panel of 
Fig.\,\ref{fig1} using $E_J$ and $f$ as adjustable parameters.  However, if we vary 
$\omega$, in addition to $\theta$, the original Eq.\,\ref{Koshelev} is inadequate.  We 
find that $f$ should have a field dependence of the form 
\begin{equation}
f(B) = g(B \sin\theta)^\alpha,
\label{fB}
\end{equation}
with $\alpha$ a small, positive exponent and $g$ a constant.

This additional $B$ dependence is immediately apparent in the limit $\theta = 90^\circ$ 
($\bf H\parallel c$).  Our measurements show that $\omega_p^2$ varies as $B^{-0.8}$, 
whereas Eq.\,\ref{Koshelev} predicts $\omega_p^2 \sim E_Jfa^2/(2k_BT)$.  This implies 
that $f\sim B_z^{0.2}$, as in Eq.\,\ref{fB} with $\alpha\simeq 0.2$.

The finite value of the anomalous exponent $\alpha$ is apparent at other tilt angles as 
well. To bring this out, it is preferable to plot, in place of $\omega_p^2$, the quantity 
$W(B,\theta) \equiv (\omega_p/2\pi)^2 B\sin\theta,$ in which the trivial dependence on 
$B_z$ is removed. We show in Fig.\,\ref{fig3} fits of $W(B,\theta)$ to the expression
\begin{equation}
W(B,\theta) =  M (B \sin\theta)^\alpha 
\exp \left[- \frac {g(B \sin\theta)^\alpha
\pi s^2 B (\cos\theta)^2}{\phi_0\sin\theta} \right].
\label{Wfit}
\end{equation}
The fitted data (54 points) are restricted to angles between 0.8$^{\circ} <\theta < 
24^{\circ}$, and frequencies between 40 and 60 GHz.  From a least-squares fit, we find at 
30 K, $\alpha$ = 0.25, $M$ = 1.31 $\times 10^{20}$, and $g$ = 1.52 ($M$ and $g$ are in SI 
units). The corresponding numbers at 50 K are 0.19, 6.52 $\times 10^{20}$, and 1.44.  We 
remark that the dependence of $W$ on $B$ and $\theta$ is non-trivial.  Nonetheless, the 
expression Eq.\,\ref{Wfit} successfully describes its behavior with only 3 adjustable
constants at each $T$. (The curves would be simple exponential decays if $\alpha$ were 
zero).  Considering the large range of $\theta$ covered, we conclude that the fit 
provides strong support for the modified Koshelev expression, with $f(B)$ given by 
Eq.\,\ref{fB}.

The quantity $f(B)$ is the number of vortices inside a correlated area.  The experiment 
shows that, as $B_z$ increases, $f(B)$ increases slightly faster than $a^{-2}$, which 
implies that the number of vortices inside a correlated area increases slowly with field.  
This is in accord with the physical expectation that correlation is relatively stronger 
at denser packing. The fits provide a quantitative description of the vortex liquid state 
at 30 and 50 K.  In terms of $M$ and $g$, the Josephson energy is given by $E_J^2 = 2k_BT 
(M\phi_0/g) (\epsilon_0 \epsilon_r/s)$, where $\epsilon_0$ and $\epsilon_r$ are the 
vacuum permittivity and dielectric constant, respectively. With $\epsilon_r\sim 25$, we 
obtain the value $E_J$ = 1.50 and 1.37 $\times 10^{-8} {\rm J/m}^2$ at 30 and 50 K, 
respectively.  These values correspond to zero-field critical current densities $J_{c0} = 
2\pi E_J/\phi_0$ = 4,600 and 4,160 A/cm$^2$.  Between 50 and 30 K, the small parameter 
$\beta E_Ja^2$ nearly doubles from 0.041 to 0.075 (in a perpendicular field $B_z$ = 1 T).  
Nevertheless, the perturbation expansion remains valid.  With $f$ given by $gB_z^\alpha$, 
we calculate $\langle \cos \phi_{n,n+1} \rangle = 0.050$ and $0.028$ at 30 and 50 K, 
respectively, in a field of 1 T.  

The picture that emerges is that, at $B_z$ = 1 Tesla, an average of 1.5 (1.4) pancakes is 
captured in a correlated patch at 30 (50) K.  These numbers increase slowly with field
as $B^\alpha$, implying that the correlation length measured in units of $a$ increases 
slowly as the packing density in the liquid increases.  The smallness of $\langle 
\cos \phi_{n,n+1} \rangle$ confirms that adjacent layers are only weakly coupled. In a 
perpendicular field, this number is further reduced as a power law $B_z^{-0.8}$. 

We briefly discuss the large-$q$ (small distance) limit. As is apparent in 
Fig.\,\ref{fig1}, the fits deviate significantly from the measurements below 1$^\circ$. 
In Eq.\,\ref{Wfit}, $\omega_p$ goes to zero rapidly as $\theta\rightarrow 0$, whereas the 
measured $B_0$ decreases linearly to a finite value.  Using our empirical power-law
description, we may express $\omega_p^2$ as $B_y^{-2(1-\eta)}$, where the exponent 
$\eta\sim 0.1$ is again fractional.  Our current understanding is that at short distances 
($r<\xi$), the Gaussian approximation for $S_1$ is not valid.  The correlation 
function seems to increase much faster than in the Gaussian expression.  This point 
requires more investigation.

In summary, we have carried out a detailed investigation of how the Josephson plasma 
resonance field $B_0$ varies with tilt angle and microwave frequency at 30 and 50 K.  At 
fixed $\omega$, Koshelev's expression \cite{Koshelev} describes the angular dependence 
remarkably well for $\theta$ larger than 1$^{\circ}$.  By introducing a weak field 
dependence to the quantity $f$, which counts the number of pancakes inside a correlated
patch, we are also able to account for the field dependence of $\omega_p$.  In this 
experiment a large number of independent measurements have been used to pin down the 
parameters $f$ and $E_J$. Thus, it yields a rather reliable estimate of these 
quantities, which characterize, respectively, the intraplane and interplane behavior of 
the vortex liquid state. Our results also relate quite naturally the anomalous exponent 
$1- \mu$ to the growth of the correlated area with $B_z$.  

We are grateful to Lev Bulaevskii, David Huse, Bernhard Keimer, Yuji Matsuda, Atsutaka 
Maeda, and M. Tachiki for many helpful discussions. The research at Princeton University 
is supported by the U.S. Office of Naval Research (Contract N00014-90-J-1013). SPB and 
NPO are also supported by funds from a MRSEC grant from the U.S. National Science 
Foundation (Grant DMR 98-09483).


%
%

\begin{figure}
\caption{(Main panel) The dependence of the resonance field $B_0$ on tilt angle $\theta$ 
and microwave frequency $\omega$ in Bi 2212 at 50 K ($\theta$ is the angle between $\bf 
H$ and the $ab$ plane). At each value of $\theta$ and $\omega$, $B_0$ is determined by 
slowly scanning $H$.  As $\theta$ decreases, $B_0$ rises to a maximum near $ 1^\circ$.  
Solid lines are fits of data above $ 1^\circ$ to Eq.\,\ref{Wfit}. In the `scattering' 
language, the region $\theta > (<) \;1^\circ$ corresponds to small (large) $q\xi$. Broken 
lines are continuations of fits to angles below 1$^\circ$. The inset shows absorption 
curves versus $H$ at 50 K at the angles indicated ($\omega/2\pi = 48$ GHz). Arrows 
indicate scan direction. }
\label{fig1}
\end{figure}

\begin{figure}
\caption{The field dependence of the plasma resonance frequency in Bi 2212 at selected 
values of $\theta$ measured at 30 K. Solid lines are fits to Eq.\,\ref{Wfit}. Data at 50 
K are similar. }
\label{fig2}
\end{figure}

\begin{figure}
\caption{The field dependence of $W\equiv (\omega_p/2\pi)^2 B\sin\theta$ at selected 
values of tilt angle at 30 K (upper panel) and 50 K (lower).  The angular dependence 
above $1^\circ$ in Fig.\,1 is dominated by $(B_0 \sin\theta)^{-1}$.  By removing this 
factor in $W$, we isolate the residual effects of the anomalous exponent $\alpha$ (all 
the curves would decrease exponentially with $B_0$ if $\alpha = 0$).  The solid lines are 
fits to Eq.\,\ref{Wfit}.  At 30 K (50 K), $\alpha = 0.25 \;(0.19)$, $M = 1.31 \;(6.52) 
\times 10^{20}$, and $g = 1.52 \;(1.44) $ in SI units.  }
\label{fig3}
\end{figure}


\begin{references}
\bibitem[*]{byline} Present address: Department of Physics, Hong Kong Univ. Sci. and 
Tech., Clear Water Bay, Kowloon, Hong Kong, China.
\bibitem{Uchida} K. Tamasaku, Y. Nakamura, and S. Uchida, Phys.\ Rev.\ Lett. {\bf 69}, 
1455 (1992).
\bibitem{Tsui1} O.\,K.\,C. Tsui \emph{et al.} , Phys.\ Rev.\ Lett. {\bf 73}, 724 (1994).
\bibitem{spie} O.\,K.\,C. Tsui, Proceedings (SPIE) Symposium on Oxide Superconductor 
Physics and Nano-Engineering II, 443 (1996).
\bibitem{Yuji1} Y. Matsuda \emph{et al.}, Phys.\ Rev.\ Lett. {\bf 75}, 4512 (1995).
\bibitem{Tsui2} O.\,K.\,C. Tsui, N.\,P. Ong, and J.\,B. Peterson, Phys.\ Rev.\ Lett. {\bf 
76}, 819 (1996).
\bibitem{Yuji2} Y. Matsuda \emph{et al.}, Phys.\ Rev.\ Lett. {\bf 78}, 1972 (1997).
\bibitem{Maeda} T. Hanaguri \emph{et al.}, Phys.\ Rev.\ Lett. {\bf 78}, 3177 (1997).
\bibitem{Sonin} E.\,B. Sonin, Phys.\ Rev.\ Lett. {\bf 79}, 3732 (1997).
\bibitem{bmt} L.\,N. Bulaevskii, M.\,P. Maley, and M. Tachiki, Phys.\ Rev.\ Lett. {\bf 
74}, 801 (1995).
\bibitem{Bula1} L.\,N. Bulaevskii, M.\,P. Maley, H. Safar, and D. Dom'inguez, Phys.\ 
Rev.\ B {\bf 53}, 6634 (1996); L.\,N. Bulaevskii, V.\,L. Pokrovsky, and M.\,P. Maley, 
Phys.\ Rev.\ Lett. {\bf 76}, 1719 (1996).
\bibitem{Koshelev} A.\,E. Koshelev, Phys.\ Rev.\ Lett. {\bf 77}, 3901 (1996).
\bibitem{Anderson} P.\,W. Anderson, in \emph{Lectures on the Many-Body Problem}, edited 
by E.\,R. Caianiello (Academic Press, New York, 1964), Vol. 2, p. 113. 
%
\end{references}
\end{document}